%% file: main.tex
\documentclass{article}
\usepackage{graphicx} 
\usepackage[colorlinks=true, linkcolor=blue, urlcolor=blue, citecolor=blue, pdftex]{hyperref}
\usepackage{amssymb} 
\usepackage{amsmath} 
\usepackage{makecell} 
\usepackage[margin=1in]{geometry}
\usepackage{array}
\usepackage{ragged2e} 
\usepackage{changes}
\usepackage{graphicx}

\definechangesauthor{Nate}
\graphicspath{ {./images/} }

\title{Sensitivity Analysis White Paper}
\author{Nate Bade \\ Lindsay Erickson}
\date{Mobius Logic}

\begin{document}

\maketitle

\section{Introduction}

\input{introduction}

\section{Decision Support Analysis Overview}

\input{decision}

\section{Literature Review and Taxonomy}

\input{lit_review_taxonomy}

\section{Methodology and Choices}

\input{pipeline}

\section{Conclusions}

\input{conclusion}

\bibliographystyle{unsrt} 
\bibliography{references} 

\end{document}

%% file: introduction.tex
Sensitivity analysis (SA) plays a critical role in simulation-based studies by quantifying how variations in input parameters influence key outputs. In defense contexts, its importance is well-documented. For example, SA has been applied to combat simulations to evaluate land combat vehicle configurations using fractional factorial designs \cite{box19612} and regression-based approaches, enabling analysts to identify influential parameters and improve system performance \cite{chau2017using}. Similarly, studies on air combat mission support effectiveness have demonstrated how simulation-based SA can validate system-of-systems performance under varying operational conditions \cite{ding2025simulation}.

Beyond individual systems, SA has supported campaign-level operations analysis, as illustrated in lessons learned from Afghanistan and Iraq, where modeling and simulation informed resource allocation and operational planning \cite{connable2014modeling}. Statistical approaches such as regression-based ranking sensitivity have also been employed in combat simulation analytics to prioritize factors affecting mission success \cite{gill2018combat}. Historical work, such as the Janus(A) combat simulation study, underscores the long-standing relevance of SA in force-on-force training scenarios \cite{feil1991sensitivity}. 

Despite these successes, applying SA to modern military exercises introduces unique challenges: high-dimensional parameter spaces, nonlinear interactions, and operational constraints that demand both computational efficiency and interpretability. These complexities necessitate automated, hybrid approaches — such as combining Morris screening for preliminary ranking with Sobol indices for detailed interaction analysis—to support decision frameworks for mission planning and resource allocation. 

%% file: decision.tex

In the decision support analysis setting, we assume we have access to a black-box model $f(X)$ depending stochastically on parameters $x \in \mathbb{R}^n$. We can sample this model $y \sim f(x)$ at points $x$ by performing an experiment or running a computation. Examples of such models might include data from a series of flight tests mimicking an operation, samples from a physics based simulation of aerodynamic operations, or runs of an agent based model with individual AIs.  

The dimensions $X^i$ may contain both controls and parameters:

\begin{itemize}
\item Parameters are dimensions of the model space a human operator has no control over, but may profit from understanding the effects of or spending research time, money, or computational cycles to determine the value of, e.g. the weather at flight time, material elasticity parameters for a particular wing design, or initialization conditions for a model.  
\item Controls are dimensions over which the operator would like to determine the effect of exerting change to, possibly at a cost. For example, a particular flight path as a series of way points, the angle variance of a wind under simulation, or particular policies for each agent in a simulation.  
\end{itemize}

Decision support is the process of understanding such a model and using it to provide actionable insights to human operators. This process involves answering a large variety of domain specific questions, but there a number of common themes and techniques that have been developed over the years. We break the decision support pipeline into five overlapping areas of techniques and inquiries. 

\textbf{Uncertainty Analysis:} How do we understand the interaction between the indeterminacy in our variables and uncertainty in likely outcomes? 

\textbf{Sensitivity Analysis:} Which factors $X^i$ contribute most to the variance in the output of our model, both locally and globally? 

\textbf{Uncertainty Quantification:} How do we characterize the distribution of model outputs, including characterizing irreducible noise? 

\textbf{Optimization:} How do we find the optimal values of quantities derived from our model given an adjustment of our controls? 

\textbf{Landscape Analysis:} How do we characterize broad regions of the domain of $X$ in terms of properties of interest in $f(X)$? I.e., where  is $f$ broadly stable, under what parameter conditions does $f$ vary above certain thresholds? What are the pathways to move from a particular sample in one region $x_1 \in \mathbb{R}_1 \subset \mathbb{R}^n$ to another sample $x_2 \in \mathbb{R}_2 \subset \mathbb{R}^n$, possibly given some path constraint?  

\textbf{Sampling Theory:} In this setting we assume that there is a cost to sampling these functions. While that cost may be low, for high dimensional or particularly noisy models the method of sampling is determinative of the success of any analysis.  

In this white paper we will focus on Sensitivity Analysis (SA), with applications to building actionable pipelines for decision support analysis. The Literature Review and Taxonomy section will explore common SA techniques, including applicability settings and a comparison of the costs and benefits associated with each method. The Methodology and Choices section will then focus on the pipeline we have developed for using sensitivity analysis to extract actionable results from agent-based models.

%% file: lit_review_taxonomy.tex
Sensitivity analysis is notoriously ambiguously defined. Indeed, according to Reed et al., ``Out of the several definitions for sensitivity analysis presented in the literature, the most widely used has been proposed by Saltelli et al. \cite{saltelli2004sensitivity} as `the study of how uncertainty in the output of a model (numerical or otherwise) can be apportioned to different sources of uncertainty in the model input,'" \cite{reed2022addressing}. Even Saltelli acknowledges that ``The term sensitivity analysis is variously interpreted in different technical communities, and problem settings. Thus a more precise definition of the terms demands that the output of the analysis be specified" \cite{saltelli2004sensitivity}. It is our goal in this literature review to precisely define what we mean by sensitivity analysis in our context. 

\begin{figure}[ht!]
    \centering
    \includegraphics[width=0.9\textwidth]{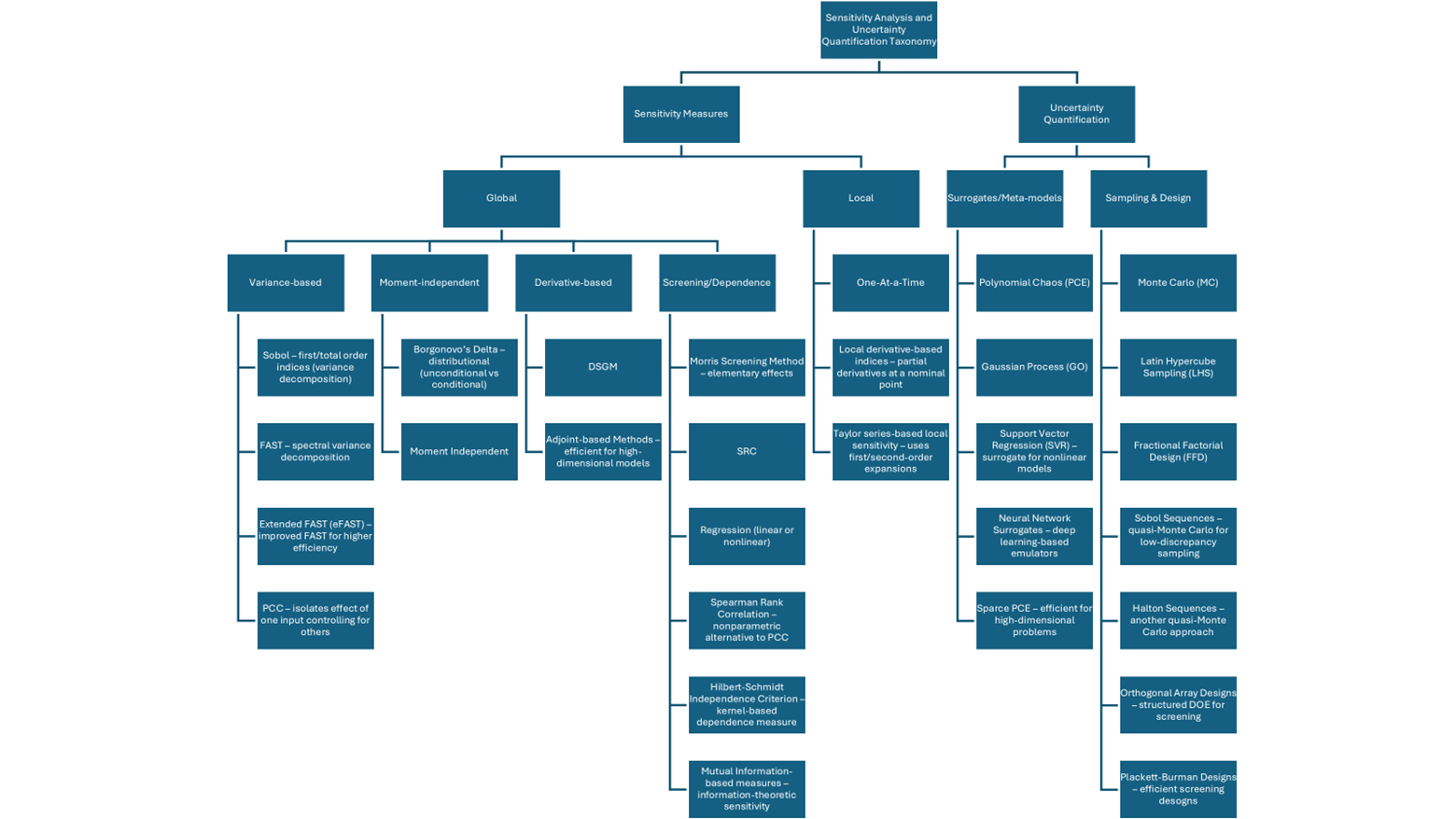}
    \caption{A taxonomy of sensitivity analysis topics.}
    \label{fig:SATaxonomy}
\end{figure}

Historically, sensitivity analysis methods have been grouped into two broad categories: local and global. Local sensitivity analysis focuses on how small perturbations in input parameters around a reference point affect model outputs, making it useful for assessing robustness near nominal conditions \cite{reed2022addressing}. A common subset of these methods additionally assumes differentially, and attempts to use local linearization to extract information. In contrast, global sensitivity analysis explores the entire feasible input space, capturing nonlinearities and interactions among factors, and it does not reference a specific point of local testing. This approach provides a more comprehensive understanding of parameter influence across an entire input space, and it is generally preferred for complex models where local linearity cannot be assumed \cite{saltelli2002making}, \cite{reed2022addressing}. Sensitivity analysis broadly breaks down into two families of related techniques. See Figure \ref{fig:SATaxonomy} for a taxonomy of sensitivity analysis topics.

\textbf{Global Sensitivity Analysis:}
Global approaches explore the entire input space using by iteratively sampling the space to explore important regions, or by decomposing global statistics from a single large sample. Global methods generally fall into a few well-defined categories and many methods specific to particular applications.  

\textbf{Local Sensitivity Analysis:}
Local techniques, such as one-at-a-time perturbations and derivative-based indices, examine behavior near a nominal point. These methods tend to be computationally efficient but limited in their ability to capture nonlinearities and interactions among inputs especially as we move away from the nominal point \cite{norton2015introduction}. Applications include understanding the effects of moving away from an observed system state.  
 
\subsection{Sampling Strategies}
Sampling strategies, including Monte Carlo, Latin Hypercube, and fractional factorial designs, remain foundational for balancing coverage and computational feasibility \cite{novak2021simulation}. Most sensitivity analysis methods rely on a particular sampling method, or family of sampling methods. We plan to dedicate another survey to sampling theory, and how different sampling methods are used across decision support theory.  

Sampling and design of experimental methods underpin both sensitivity and uncertainty analyses by defining input sets for simulation runs. Morris Screening, Monte Carlo methods, and Latin Hypercube Sampling remain foundational techniques for uncertainty propagation \cite{mckay2000comparison}, while quasi-Monte Carlo sequences such as Sobol and Halton improve convergence through low-discrepancy sampling \cite{mckay2000comparison}. Structured designs, including orthogonal arrays, Plackett–Burman designs, and fractional factorial approaches, enable efficient screening and interaction estimation under resource constraints \cite{saltelli2000sensitivity}. 

\subsection{Uses of Sensitivity Analysis}
Within these categories, sensitivity analysis serves multiple purposes. Factor screening techniques aim to identify and rank inputs based on their qualitative impact on outputs, enabling analysts to eliminate parameters with negligible influence early in the modeling process \cite{siebertz2010statistische}. Local sensitivity analysis is often employed to evaluate stability for a specific configuration, determining how small changes in factor values affect performance metrics \cite{siebertz2010statistische}. Global sensitivity analysis, on the other hand, quantifies the relative importance of parameters across their entire range, offering insight into both main effects and interactions \cite{saltelli2002making}, \cite{siebertz2010statistische}. These methods collectively support systematic exploration of model behavior under uncertainty, guiding decisions about which factors warrant detailed study and which can be safely fixed or ignored \cite{grznar2024comprehensive}.

\subsection{Methods in Sensitivity Analysis And Uncertainty Quantification} 
  
Sensitivity analysis and uncertainty quantification form the analytical backbone of mission-level modeling and simulation, providing insight into how variations in input parameters influence key performance metrics and how uncertainty propagates through complex systems. These methods can be broadly categorized into sensitivity measures and uncertainty quantification techniques, each serving distinct but complementary roles in model evaluation and decision support. 

\subsubsection{Variance Based Methods}
Variance based methods refer to a family of techniques that derive insight from understanding the global variance $V$ as $X$ ranges over the entire domain. These methods are often used to try to understand which factors of the input variables most effect the change in the output, either so those factors can be accurately determine to reduce uncertainty, to drop factors that are largely irrelevant to outcomes, or to localize on a maximally uncorrelated set of important factors.

The main tool is variance decomposition into Sobol indices. Sobol indices decompose variance $V$ into contributions from individual inputs $x_i$ (i.e., first order effects) and their interactions $x_i \ldots x_j$ (higher order effects) \cite{sobol2001global}, offering interpretable metrics for ranking factors by main and total effects, especially when inputs are understood to be orthogonal. 

For non-orthogonal inputs, partial correlation coefficients methods isolate the effect of one input while controlling for others \cite{saltelli2000sensitivity}. Techniques like FAST and its extended variant eFAST employ spectral decomposition to estimate the variance contributions at different orders efficiently \cite{saltelli2000sensitivity}, \cite{ryan2018fast}.  

\subsubsection{Moment Independent Methods}
Generalizing variance-based approaches, moment-independent measures such as Borgonovo’s Delta assess sensitivity through distributional differences between unconditional and conditional outputs \cite{borgonovo2007new}, providing robustness in cases where variance alone is insufficient. Given some notation of distributional distance $d$, the sensitivity measures are defined as $$| \xi = E_{x_i}[d(F(x), F(x | x_i))] |.$$


Moment independent measure capture information about the entire distribution, and may work better for distributions that are not known to be dominated by a particular moment.  

\subsubsection{Derivative-based methods}
If derivatives are available globally, they can be a powerful tool for local or global sensitivity analysis, giving rich information about functional change around any given point. Methods such as autodiff and adjoint modeling can often be used to lower the computational cost of derivative modeling, however full global information can rapidly become computationally expensive. These methods are particularly effective for physics-based simulations \cite{saltelli2000sensitivity}.  

\subsubsection{Screening}
Screening methods attempt to cheaply rank input dimensions by importance, without necessarily providing measures of the absolute importance. They are often used in situations where precisely fixing variables may be expensive, and so only a small number of variables can be fixed by either experimental runs or analytic efforts. Due to the assumptions of sampling restriction, screening methods are deeply related to efficient sampling theory, often using simple measures.  

For example, Morse Screening uses short random walks of finite differences in the cardinal directions $e_i$ to efficiently sample the global effects of changing each input factor $x_i$. The means of these global effects are then used to estimate relative importance.  

Additional methods include standardized regression coefficients, Spearman rank correlations \cite{morris1991factorial}, and kernel-based metrics like the Hilbert–Schmidt Independence Criterion, provide cost-effective ways to detect influential factors and nonlinear dependencies early in the analysis process \cite{gretton2005measuring}.  

\subsubsection{Local Methods}
Local sensitivity analysis complements global approaches by focusing on behavior near a nominal operating point. One-at-a-time perturbations, derivative-based indices, and Taylor series expansions provide insight into local robustness and directional effects, supporting calibration and control in operational regimes where small changes can have significant consequences \cite{saltelli2000sensitivity}.  

Many local methods can be integrated out to global methods using proper sampling theory; however, often local methods are considered local because they become prohibitively computationally expensive outside of a region around a normal operating point.  

\subsubsection{Uncertainty Quantification}
Uncertainty quantification techniques address the propagation and characterization of uncertainty in model outputs, enabling risk-informed decision-making and confidence assessment. Uncertainty quantification contrasts with sensitivity analysis in that it is interested in fully characterizing the stochastic nature of a model, both as a function of it’s inputs $x$ and as irreducible noise, as opposed to understanding the importance of particular variables.  

For example, ``What is the range of the model compared to the range of the irreducible noise in the model?" would be a question of uncertainty quantification. Whereas, ``which factor $x_i$ contributes the most globally to the change in model?" would be a sensitivity analysis question.  

Surrogate modeling plays a central role in this domain, offering computationally efficient approximations of expensive simulations. The simplest form is using universal function approximators, such as spline fitting or neural networks, to fit an underlying model with an additional noise term.  

Probability-based surrogate methods include using Polynomial Chaos Expansion to represent stochastic outputs through orthogonal polynomial bases \cite{mara2021polynomial}, while sparse variants improve scalability in high-dimensional settings. Gaussian process emulators provide probabilistic predictions with uncertainty bounds \cite{williams2006gauss}, facilitating adaptive sampling and sequential design. Other surrogate approaches, including support vector regression and neural network-based models, offer flexibility for nonlinear and high-dimensional relationships, with neural networks particularly suited for large-scale emulation tasks \cite{williams2006gauss}.  

To guide practitioners, recent reviews synthesize global versus local method selection and provide actionable frameworks for scaling SA in complex models \cite{francom2025review}, \cite{norton2015introduction}. Toolchains combining LHS, Sobol indices, and surrogate modeling offer pragmatic solutions for computationally intensive simulations \cite{novak2021simulation}. 
 
The table below compares common SA methods by computational cost, interpretability, robustness, and typical use cases. 
 
Comparison Table: Sensitivity Analysis Methods vs. Characteristics 
\begin{table}[h!]
\centering
\small
\setlength{\tabcolsep}{4pt} 
\renewcommand{\arraystretch}{1.1} 
\begin{tabular}{p{2.2cm} p{2.0cm} p{2.0cm} p{2.0cm} p{1.8cm} p{3.0cm}}
\hline
\textbf{Method} &
\textbf{Computa- tional Cost} &
\textbf{Interpret- ability} &
\textbf{Captures Interactions} &
\textbf{Robustness} &
\textbf{Typical Use} \\
\hline
OAT &
Low &
High &
No &
Low &
Sanity checks \\
Morris &
Moderate &
Moderate &
Partial &
Partial &
High-dimensional screening \\
Sobol &
High &
High &
Yes &
High &
Rigorous global sensitivity analysis \\
FAST &
High &
Moderate– High &
Yes &
High &
Efficient variance- based sensitivity analysis \\
Regression &
Low &
Moderate &
Limited &
Limited &
Quick ranking \\
DGSM &
Moderate &
Moderate &
Indirect &
Requires smoothness &
Screening and bounding \\
Moment-independent &
High &
Moderate &
Yes &
High &
Non-monotonic responses \\
\hline
\end{tabular}
\end{table}

\subsection{Settings For Sensitivity Analysis} 
There are four main settings for sensitivity analysis. The first setting, factor prioritization, seeks to identify the most influential input variable—defined as the factor whose precise knowledge would yield the greatest reduction in output variance. This approach enables rational decision-making under uncertainty because determining the most important factor implies that its true value can be established through targeted measurement or research. Factor prioritization typically assumes that inputs are fixed one at a time, which limits the detection of interactions as a natural consequence of the method. In practice, this setting is particularly useful for guiding experimental design and resource allocation, as it informs which parameters warrant more accurate estimation or tighter control. 

The second setting, factor fixing, also referred to as screening, aims to identify inputs that can be held constant without significantly affecting the variability of model outputs. This approach is often employed to simplify models by removing non-influential factors, thereby reducing computational burden while preserving fidelity. Factor fixing can also serve as a diagnostic tool to validate or challenge a given model representation. Techniques such as the Morris method and variance-based approaches are commonly used in this setting, with the necessary and sufficient condition for a factor to be considered non-influential expressed as the expectation of the conditional variance being zero. 

The third setting, variance cutting, is particularly relevant in risk assessment contexts. Its objective is to reduce the unconditional variance of the output to a predefined threshold by fixing the smallest possible number of factors. Rather than seeking an optimal solution, this setting supports informed choices about which parameters to constrain in order to achieve acceptable levels of uncertainty. Variance cutting can be addressed through conditional variance analysis, providing a structured way to manage risk in complex systems. 

Finally, the factor mapping setting focuses on understanding how combinations of input values correspond to specific regions of the output space. This approach often employs Monte Carlo simulations to classify outputs into acceptable and unacceptable categories based on empirical distributions or predefined thresholds. By filtering simulation results and applying statistical tests such as the Smirnov test, analysts can trace back from output classifications to the input factors most responsible for driving these distinctions. Factor mapping thus answers the critical question of which variables govern the likelihood of achieving desired performance outcomes, making it a powerful tool for design optimization and operational planning. 


\subsection{Challenges With Sensitivity Analysis}
A recurring challenge in SA is distinguishing correlation from causation. Variance-based indices apportion output variance to inputs under a sampling model but do not establish causal direction without additional assumptions or experimental designs \cite{norton2015introduction}, \cite{grznar2024comprehensive}, \cite{francom2025review}. In defense simulations, mission phasing alone cannot establish causality because phases are not isolated; they share resources, constraints, and parameter influences. Behaviors in one phase often depend on prior conditions and latent factors that span multiple phases—for example, detection success in an early phase can influence engagement strategies later, while parameters such as communication latency or sensor range affect multiple phases simultaneously. These interdependencies mean that temporal sequence does not imply causal independence. Sensitivity analysis can rank influential parameters within phases, but causal claims require explicit causal modeling or controlled interventions \cite{chau2017using}, \cite{gill2018combat}. 
 
Computational complexity is another concern. Global SA methods often require large sample sizes, creating tension between exhaustive variance decomposition and high-level screening. Pragmatic strategies include two-stage SA (Morris for screening followed by Sobol or FAST for detailed analysis) \cite{francom2025review}, \cite{novak2021simulation}, surrogate modeling using polynomial chaos or Gaussian processes \cite{novak2021simulation}, \cite{norton2015introduction}, and experimental designs such as fractional factorials or LHS for efficient sampling \cite{chau2017using}, \cite{gill2018combat}. Moment-independent metrics are recommended when variance-based views fail to capture distributional changes critical to mission outcomes \cite{grznar2024comprehensive}. 

Few defense SA publications explicitly address causality in complex, phased missions, highlighting a gap in methodologies that integrate computational tractability with causal inference. Our approach will clarify the boundary between influence and causality and propose workflows combining screening, surrogate-based global SA, and cause-effect reasoning. 
 
Saltelli and colleagues have been instrumental in formalizing global sensitivity analysis, particularly variance-based methods and their practical implementation. Their works, such as Sensitivity Analysis in Practice and Global Sensitivity Analysis: The Primer, provide comprehensive guidelines for applying Sobol indices, FAST, and screening techniques in complex models \cite{saltelli2004sensitivity}, \cite{saltelli2008global}. Kleijnen’s contributions focus on the design and analysis of simulation experiments, bridging classical Design of Experiments with modern approaches like LHS and Kriging metamodels. His tutorials and books, including Design and Analysis of Simulation Experiments, remain foundational for practitioners seeking efficient and statistically sound SA in simulation contexts \cite{kleijnen2015design}, \cite{kleijnen2005overview}.

%% file: pipeline.tex
We now outline the methodological framework adopted for sensitivity analysis in the context of military exercise simulations. Our approach combines a two-stage sensitivity analysis process with additional techniques for computational efficiency and robustness. 
 
The first stage employs the Morris method for screening. Morris, based on elementary effects, provides a qualitative ranking of parameters and flags potential nonlinearity and interaction effects. This method is particularly suited for high-dimensional parameter spaces, enabling us to identify influential factors early without incurring prohibitive computational costs \cite{saltelli2004sensitivity}, \cite{saltelli2008global}. 
 
The second stage applies Sobol indices for global sensitivity analysis. Sobol’s variance-based decomposition quantifies both first-order and higher-order effects, offering rigorous insight into parameter interactions. This level of detail is essential for mission planning, where understanding parameter interplay can inform Course of Action (COA) comparisons and resource allocation \cite{saltelli2004sensitivity}, \cite{saltelli2008global}. 
 
To enhance efficiency, we incorporate clustering techniques to group parameter regions with similar sensitivity profiles and adaptive resampling in areas of high gradient change. These additions allow us to refine sensitivity estimates without exhaustive sampling, aligning with operational constraints and computational limits.  
 
Integration into the simulation workflow follows a structured pipeline: parameter sampling, simulation execution, Morris screening, Sobol analysis, clustering and adaptive resampling, and finally, decision support. This workflow ensures that sensitivity insights directly inform mission evaluation and planning. 
 
Assumptions underlying this methodology include parameter independence for sampling, fidelity of the simulation model to operational realities, and predefined parameter ranges based on expert input. These assumptions satisfy the requirements of the selected methods and ensure interpretability of results. 
 
Our choices reflect the mission objective of enabling decision-making across varied configurations of fixed and flexible parameters. By combining screening, global analysis, and adaptive techniques, we provide a robust framework for analyzing complex military scenarios under uncertainty. 
 
Workflow Diagram: Two-Stage Sensitivity Analysis with Clustering and Adaptive Resampling 

\begin{table}[h!]
\centering
\small
\setlength{\tabcolsep}{4pt}
\renewcommand{\arraystretch}{1.1}
\begin{tabular}{
    >{\RaggedRight\arraybackslash}p{3.0cm} 
    >{\RaggedRight\arraybackslash}p{2.8cm} 
    >{\RaggedRight\arraybackslash}p{3.6cm} 
    >{\RaggedRight\arraybackslash}p{2.0cm} 
    >{\RaggedRight\arraybackslash}p{3.6cm} 
}
\hline
\textbf{Pipeline} &
\textbf{Flow} &
\textbf{Purpose} &
\textbf{Computa- tional Cost} &
\textbf{Typical Use Case} \\
\hline
Pipeline 1: Screening → Variance Decomposition &
Morris followed by Sobol or FAST &
Identify influential factors quickly, then quantify detailed variance contributions &
Moderate to High &
High-dimensional models where full Sobol analysis is too expensive initially \\
Pipeline 2: Surrogate-Assisted SA &
Polynomial Chaos or Gaussian Process → Sobol or Borgonovo’s Delta &
Reduce cost by approximating expensive simulations with surrogate models &
Moderate (after surrogate fit) &
Complex, computationally expensive models (e.g., physics-based simulations) \\
Pipeline 3: Distribution-Sensitive Analysis &
Moment-independent metrics like Borgonovo’s Delta &
Capture effects beyond variance, focusing on full output distribution &
High &
Mission-critical scenarios where tail behavior or risk metrics matter \\
Pipeline 4: Sampling-Driven Design &
Latin Hypercube Sampling or Fractional Factorial → Regression/Correlation methods like SRC, PCC &
Efficiently explore input space for ranking under near-linear assumptions &
Low &
Early-stage analysis or models with approximately linear relationships \\
\hline
\end{tabular}
\end{table}

\subsection{Application to Military Simulation}
Sensitivity analysis plays a critical role in military simulation by enabling decision-makers to evaluate mission success under varying operational conditions and parameter configurations. The overarching objective is to identify the factors that most influence mission outcomes and to optimize resource allocation in complex, uncertain environments. 

Modern electronic warfare systems operate in highly dynamic settings, making it challenging to replicate real-world conditions during testing. Programs often rely on multiple test environments—such as anechoic chambers and over-the-air facilities—to capture diverse operational scenarios. To maximize efficiency across these sites, design of experiments principles are essential, allowing flexible prioritization as new insights emerge during testing. Effective communication throughout planning, execution, and analysis is equally important, as errors can arise from numerous sources, including test setup, human factors, and system malfunctions. Rapid identification of these sources enables the testing community to address anomalies while maintaining budget and schedule constraints \cite{destefan2022ewar}. 

The complexity of adversary systems further underscores the need for robust analytical methods. Modern sensors and command-and-control architectures are increasingly agile, employing adaptive waveforms and cognitive techniques to counter electronic attack. These capabilities demand that friendly systems become more resilient and adaptive to maintain operational effectiveness in contested electromagnetic environments \cite{destefan2022ewar}. 

Constructive military simulations must also account for organizational context. Lower echelons of the military hierarchy focus on operational details such as logistics and resupply, requiring models that capture fine-grained aspects of mission execution. For example, unit-level air operations may necessitate detailed terrain modeling for mission rehearsal. In contrast, higher-level decision-making emphasizes broader trade-offs and strategic outcomes, which call for aggregated representations of mission dynamics \cite{hill2001applications}. 

The simulation framework developed for this study models a representative mission scenario comprising multiple phases, including detection, engagement, and sustainment. Key parameters include sensor range, communication latency, platform speed, weapon system effectiveness, and environmental conditions such as wind and precipitation. Success metrics are defined in terms of detection probability, engagement success rate, mission completion time, and resource utilization efficiency \cite{dahmann2024mission}. These metrics provide a basis for evaluating mission performance under uncertainty and for identifying leverage points for improvement. 

Sensitivity analysis integrates into this workflow as a decision-support capability. The process begins with parameter sampling across predefined ranges, followed by execution of simulation runs. Morris screening is applied to identify influential parameters, and Sobol’ analysis quantifies main and interaction effects. Clustering and adaptive resampling refine sensitivity estimates in regions of high variability. Insights from these analyses feed into a decision framework aligned with the DoD Mission Engineering Guide 2.0, which emphasizes rigorous analysis for mission assurance and adaptability \cite{saltelli2008global}. Example outputs include screening plots highlighting parameter influence and ranking tables summarizing Sobol’ indices. These artifacts enable analysts to prioritize parameters for tuning and assess trade-offs between mission objectives and resource constraints. While mock outputs are presented here for illustration, actual implementation will generate data-driven visualizations to support operational decision-making. 

Finally, sensitivity metrics derived from this process can inform machine learning applications, such as surrogate modeling and predictive analytics, which may be explored in subsequent research \cite{kleijnen2015design}.  

The application of sensitivity analysis within mission-level simulations aligns closely with the principles outlined in the DoD Mission Engineering Guide 2.0 (MEG) \cite{dahmann2024mission}, particularly Chapter 6, which addresses mission engineering analysis. The MEG emphasizes a disciplined approach to evaluating mission performance through structured experimentation, iterative refinement, and transparent documentation. Sensitivity analysis directly supports these objectives by providing quantitative insight into the influence of key parameters on mission outcomes. 

\subsubsection{Alignment with MEG Analysis Framework}
\begin{enumerate}	
\item Design of Analysis and Run Matrix Development 
Section 6.1 of the MEG underscores the importance of defining an evaluation framework and constructing a run matrix that captures relevant mission conditions \cite{dahmann2024mission}. Sensitivity analysis complements this process by systematically identifying and prioritizing influential parameters. Techniques such as Morris screening and Sobol variance decomposition enable analysts to focus computational resources on factors that exert the greatest impact on mission success metrics, thereby improving the efficiency and relevance of the run matrix. 

\item Execution and Iterative Refinement 
The MEG advocates for iterative execution of baseline and alternative mission configurations to validate fidelity and explore trade spaces. Sensitivity analysis operationalizes this guidance by quantifying parameter effects and uncertainty propagation, informing adjustments to simulation scenarios. This iterative feedback loop ensures that mission models remain robust and decision-relevant. 

\item Transparency and Traceability 
A core principle of the MEG is the documentation of assumptions, constraints, and analytical methods to build confidence in results. Sensitivity analysis inherently supports this requirement by producing interpretable outputs, such as screening plots, ranking tables, and uncertainty visualizations, that clearly communicate the drivers of mission performance. These artifacts enhance traceability and facilitate stakeholder engagement. 

\item Decision-Making and Trade Space Exploration 
The MEG emphasizes the need for quantitative analysis to inform capability trades and investment decisions. Sensitivity analysis provides a rigorous foundation for such decisions by identifying high-leverage parameters and quantifying their impact on mission objectives. This enables decision-makers to allocate resources effectively and prioritize technologies that deliver the greatest operational benefit. 
\end{enumerate}

\subsubsection{Implications for Military Applications}
By embedding sensitivity analysis within the mission engineering process, analysts can transition from qualitative assessments to evidence-based decision-making. This approach not only aligns with MEG guidance but also enhances the ability to evaluate alternative courses of action, assess robustness under uncertainty, and support acquisition and operational planning. Furthermore, the integration of sensitivity analysis outputs into digital engineering environments ensures consistency, reusability, and scalability across mission studies. 
 
\subsection{Sensitivity Auditing}
Sensitivity auditing is a structured approach to evaluating decision relevant models that extends beyond conventional sensitivity analysis. Whereas sensitivity analysis prioritizes quantifying how input uncertainties propagate to outputs, sensitivity auditing investigates the broader inferential chain.  Sensitivity auditing focuses on framing choices, implicit assumptions, institutional settings, and the ethics of quantification, through which models become authoritative in policy debates \cite{lo2023sensitivity}, \cite{saltelli2014all}. It is designed for contexts where stakes are high, values are contested, and model outputs risk being interpreted with unwarranted confidence. The aim is not to refine a model’s internal mechanics per se, but to ensure that its deployment is transparent, accountable, and appropriately circumspect \cite{lo2023sensitivity}, \cite{saltelli2014all}. 

The rationale for sensitivity auditing arises from the recognition that models acquire epistemic authority through their complexity, precision, and presentation by technical experts, often in ways that discourage scrutiny of their assumptions and framing. This authority, if not audited, can obscure normative commitments and amplify selective interpretations that suit preexisting agendas. Sensitivity auditing therefore considers how models are constructed, documented, and communicated, with particular attention to the conditions under which uncertainty is either over  or under stated. It addresses the risk that quantification can implicitly shape policy by privileging certain metrics, thresholds, or optimization criteria, thereby influencing whom policies benefit and which tradeoffs are rendered salient \cite{lo2023sensitivity}, \cite{saltelli2014all}. 

The seven rule checklist presented by Lo Piano, et al., offers an operational playbook for auditing models intended to inform policy \cite{lo2023sensitivity}. Each rule targets a distinct class of epistemic risks and practical remedies: 

\begin{enumerate}
\item Guard Against Rhetorical Uses of Mathematics 
Models should be proportionate in complexity to their purpose and evidentiary base. Auditing asks whether mathematical sophistication is justified by available data, theory, and decision needs, and whether model scope has been stretched beyond its legitimate domain. The objective is to prevent complexity from functioning as a rhetorical device that suppresses critical engagement and inflates perceived certainty \cite{lo2023sensitivity}, \cite{saltelli2014all}. 

\item Adopt an Assumption Hunting Attitude 
Sensitivity auditing requires systematic elicitation of both explicit and implicit assumptions across the model’s lifecycle of conceptualization, formalization, calibration, and interpretation. This includes structural idealizations (e.g., equilibrium, representative agents), boundary conditions, and value choices about what is modeled and what is held constant. Assumption logs and pedigree assessments help prevent institutional memory loss and enable external review \cite{lo2023sensitivity}. 

\item Detect “Garbage In–Garbage Out” Dynamics 
Auditing evaluates whether uncertainty has been unduly magnified or, more commonly, understated. Understatement can produce deceptively crisp outputs that mask data limitations, identification problems, or structural uncertainty. Conversely, overstatement can paralyze decision making without clarifying which uncertainties matter. Auditors assess input data quality, measurement validity, and the alignment between uncertainty characterization and the model’s inferential claims \cite{lo2023sensitivity}, \cite{saltelli2014all}. 

\item Identify Sensitive Assumptions Early 
Before public scrutiny, modelers should perform rigorous sensitivity analysis to reveal which inputs, structural features, or modeling choices most influence results. This proactive step reduces the risk that external critics will uncover decisive vulnerabilities and ensures that communication of results appropriately prioritizes those choices and their plausible ranges \cite{lo2023sensitivity}. 

\item Aim for Transparency 
Transparency encompasses accessible documentation, intelligible code, reproducible workflows, and clear origins of data and parameters. Black box models undermine trust in policy settings; auditors evaluate whether the model can be inspected and replicated by qualified third parties, and whether explanations are written for diverse stakeholders, not only model developers \cite{lo2023sensitivity}, \cite{saltelli2014all}. 

\item Do the Right Sums, Not Just the Sums Right 
Auditing challenges the implicit neutrality of quantification by asking whether the modeled objective functions, metrics, and performance criteria faithfully represent societal concerns, including distributional impacts and potential winners and losers. It foregrounds reflexivity about how measurement choices condition policy options and public discourse, and whether analyses include perspectives that are otherwise marginalized by technocratic frames \cite{lo2023sensitivity}, \cite{saltelli2014all}. 

\item Perform State of the Art Uncertainty and Sensitivity Analyses 
Sensitivity auditing assess whether uncertainty and sensitivity analyses meet current methodological standards.  These standards include global over local approaches where appropriate, robust experimental designs, sufficient sampling, and transparent reporting of assumptions and limitations. They also examine whether results are communicated in ways that prevent over interpretation, such as by emphasizing credible intervals, robustness checks, and scenario diversity \cite{lo2023sensitivity}. 
\end{enumerate}

Sensitivity auditing is conceived as a late stage, pre-deployment safeguard. It complements model verification and validation by scrutinizing whether the model, after calibration and optimization, is ready to enter policy debates with appropriate caveats and disclosures. In this sense, auditing parallels the logic of a tax audit: it does not seek to redesign the ``accounting system" but to ensure that the products of that system can withstand external scrutiny, serve the public interest, and avoid spurious precision or false confidence \cite{saltelli2014all}. The process thereby protects both the integrity of modeling and the legitimacy of policy that relies upon it \cite{lo2023sensitivity}, \cite{saltelli2014all}.

%% file: conclusion.tex
Sensitivity auditing repositions models from instruments of definitive prediction to tools for structured deliberation under uncertainty. By making assumptions explicit, emphasizing distributional consequences, and preventing false precision, auditing enhances credibility and supports more responsible decision making. It fosters a culture in which models are judged not solely by technical virtuosity, but by transparency, robustness, and ethical awareness \cite{lo2023sensitivity}, \cite{saltelli2014all}. 

Sensitivity auditing provides a principled, operational framework for ensuring that decision relevant models are deployed with appropriate humility, clarity, and accountability. Through its seven rules, it extends the remit of model evaluation to include the social and ethical dimensions of quantification, thereby strengthening both the credibility of modeling and the legitimacy of the policies it seeks to inform \cite{lo2023sensitivity}, \cite{saltelli2014all}.